\documentclass[aps,prl,twocolumn]{revtex4}

\usepackage{color}
\usepackage{times}
\usepackage{epsfig}
\usepackage{amsmath}
\usepackage{amssymb}
\usepackage{amsmath, amsthm, amssymb,  mathrsfs, epsfig}
\usepackage{dcpic, pictex, pinlabel, setspace, rotating}
\usepackage{faktor}

\def\bra#1{\left\langle#1\right|}
\def\ket#1{\left|#1\right\rangle}

\def\bea{\begin{eqnarray}}      \def\eea{\end{eqnarray}}
\def\Zto {\mathbb{Z}_2}

\begin{document}

\title{A new class of $(2+1)$-d topological superconductor with $\mathbb{Z}_8$ topological classification}

\author{Xiao-Liang Qi$^{1}$}
\affiliation{
$^1$Department of Physics, Stanford University, Stanford, CA 94305, USA
}

\date{\today}

\begin{abstract}
The classification of topological states of matter depends on spatial dimension and symmetry class. For non-interacting topological insulators and superconductors the topological classification is obtained systematically and nontrivial topological insulators are classified by either integer or $Z_2$. The classification of interacting topological states of matter is much more complicated and only special cases are understood. In this paper we study a new class of topological superconductors in $(2+1)$ dimensions which has time-reversal symmetry and a $\mathbb{Z}_2$ spin conservation symmetry. We demonstrate that the superconductors in this class is classified by $\mathbb{Z}_8$ when electron interaction is considered, while the classification is $\mathbb{Z}$ without interaction. 
\end{abstract}


\maketitle


Topological states of matter (TSM) are ground states of gapped quantum systems which cannot be adiabatically tuned to a topologically trivial state without going across a phase transition. The first TSM discovered in nature are integer and fractional quantum Hall states\cite{klitzing1980,tsui1982}. Many topological states have gapless surface states on the boundary which are topologically protected within the given symmetry class. For example, the integer quantum Hall states have chiral fermion edge states which propagates unidirectionally along the edge of the two-dimensional system, and is robust against arbitrary perturbation as long as the bulk remains gapped. The number of chiral channels is an integer which determines the Hall conductance and characterizes the corresponding quantum Hall state.

Since the recent discovery of time-reversal invariant topological insulators (TI)\cite{hasan2010,moore2010,qi2011RMP} topological states of matter has been understood much more systematically in generic spatial dimensions and symmetries.\cite{qi2008b,schnyder2008,kitaev2009} For gapped non-interacting fermion systems, {\it i.e.} band insulators without electron correlation and gapped superconductors in the mean-field theory sense, there are ten symmetry classes in each spatial dimension\cite{zirnbauer1996,altland1997} if we only consider symmetries which preserve the positions of electrons, such as time-reversal symmetry and particle-hole symmetry. In each spatial dimension, TSM exists in five of the ten symmetry classes, which always include three classes with integer classification, and two classes with $Z_2$ classification.\cite{schnyder2008,kitaev2009} The topological classification depends on the spatial dimension and symmetry class in a periodic way\cite{kitaev2009} which originates from the Bott periodicity. When electron interaction is considered, the problem is much more complicated and the topological classification in generic dimensions and symmetry classes has not been completely understood. Topological invariants based on single particle Green's function has been proposed for interacting TSM\cite{wang2010b,wang2012,gurarie2011,essin2011}. In one dimension (1D), the classification of gapped states has been studied systematically by making use of matrix product states and entanglement spectrum\cite{fidkowski2011,turner2011}. In particular, L. Fidkowski and A. Kitaev proposed an explicit example of one-dimensional time-reversal invariant superconductors\cite{fidkowski2010} for which the topological classification is $\mathbb{Z}$ for non-interacting states but is reduced to $\mathbb{Z}_8$ when interaction is considered. In dimensions higher than one, a systematical approach of constructing interacting symmetry-protected topological states (defined as the TSM in a certain symmetry class which becomes trivial when the symmetry is broken) in spin models has been proposed based on group cohomology\cite{chen2011b}. The relation between such states and the non-interacting TSM's in generic cases remains an open question.\footnote{After finishing this work, I notice a recent paper\cite{gu2012} which generalizes the group cohomology approach to fermion systems. Whether the result of the current paper can be understood by the approach of \cite{gu2012} is an interesting question.} 

In this paper, we propose a new class of TSM (labeled by $D'$) which in two dimensions is classified by $\mathbb{Z}_8$. 
This symmetry class is defined by two discrete symmetries, the time-reversal symmetry and an additional global $\mathbb{Z}_2$ symmetry. Physically, this symmetry class consists of time-reversal invariant superconductors of spinful fermions with the number of spin-up electrons (and also that of spin-down electrons) conserved modular $2$. Without interaction, the topological classification is $\mathbb{Z}$, the same as time-reversal breaking superconductors (class $D$) in 2D\cite{read2000}. With electron interactions, we show that the classification is reduced to $\mathbb{Z}_8$, by analyzing the edge states and topological defects. This state can be considered as a higher dimensional generalization of the $\mathbb{Z}_8$ topological superconductor (TSC) proposed in Ref. \cite{fidkowski2010}.

\noindent{\it Definition of the symmetry class and the classification of non-interaction systems}. Consider a 2D electron system with the generic Hamiltonian $H=H(c_{i\sigma},c_{i\sigma}^\dagger)$, with $\sigma=\uparrow,\downarrow$ two spin components, and $i$ labels the lattice sites. 
The symmetries we consider are time-reversal symmetry defined by
\bea
T^{-1}c_{i\uparrow} T=c_{i\downarrow},~T^{-1}c_{i\downarrow }T=-c_{i\uparrow}\label{Tdef}
\eea
and the $\Zto$ spin conservation symmetry named by $R$:
\bea
R^{-1}c_{i\uparrow} R=c_{i\uparrow},~R^{-1}c_{i\downarrow} R=-c_{i\downarrow}\label{Rdef}
\eea
It should be noticed that $T$ is an anti-unitary symmetry and $R$ is a unitary symmetry. In an explicit form $R=(-1)^{N_\downarrow}$ with $N_\downarrow$ the total number of spin down electrons. An equivalent definition of $T$ and $R$ symmetries which is basis independent is
\bea
T^2&=&F,~F^2=1,~R^2=1,~T^{-1}RT=FR\nonumber\\
T^{-1}iT&=&-i,~R^{-1}iR=i
\eea
in which $F=(-1)^{N_\uparrow+N_\downarrow}$ is the fermion number parity. The second line means that $T$ is anti-unitary and $R$ is unitary. We label the symmetry class defined by symmetries $T$ and $R$ as class $D'$ due to its relation to the class $D$ of time-reversal symmetry breaking superconductors, which will be come clear in later part of the draft.

Now we consider the simple situation when the Hamiltonian is quadratic in $c_{i\sigma},c_{i\sigma}^\dagger$. Apparently, any term mixing spin up and down electrons such as $c_{i\uparrow}^\dagger c_{j\downarrow}$ and $c_{i\uparrow}c_{j\downarrow}$ is odd under the $\Zto$ symmetry $R$, such that the quadratic Hamiltonian $H_{\rm quad}$ is decoupled into a direct sum of that of the spin up and down subsystems:
\bea
H_{\rm quad}&=&H_\uparrow+H_\downarrow\nonumber\\
H_\sigma&=&\sum_{i,j}\left(c_{i\sigma}^\dagger,~c_{i\sigma}\right)h^{\sigma}_{ij}\left(\begin{array}{c}c_{i\sigma}\\c_{i\sigma}^\dagger\end{array}\right),~\sigma=\uparrow,\downarrow
\label{Hbulk}
\eea
The single-particle Hamiltonians $h_{ij}^\uparrow$ and $h_{ij}^{\downarrow}$ are related by the time-reversal invariance condition:
\bea
T^{-1}H_\uparrow T=H_\downarrow\Rightarrow {h_{ij}^\uparrow}^* =h_{ij}^\downarrow\label{Tofh}
\eea
Thus all non-interacting states in this class are direct products of spin up and down superconductor ground states, and the spin down state is determined by the spin up state due to time-reversal symmetry. Consequently the topological classification of $H_{\rm quad}$ reduces to that of the spin up subsystem $H_\uparrow$, which is a Hamiltonian of a generic 2D superconductor, {\it without} time-reversal symmetry. (The time-reversal symmetry is recovered only when the spin-down state is also included.)

It is well-known that the 2D time-reversal breaking superconductors are classified by integer, in the same way as integer quantum Hall states.\cite{read2000} The topological invariant characterizing this class of topological superconductors is the Chern number of the single-particle Hamiltonian. For a translation invariant system, if we denote $\ket{n,{\bf k}}$ as the eigenstates of the single particle Hamiltonian $h^\uparrow$, with correspondingly eigenvalues $E_n({\bf k})$, the Chern number is defined as the total flux of the Berry phase gauge field in the Brillouin zone: $C_1=\frac1{2\pi}\int d^2{\bf k}\left(\partial_xa_y-\partial_ya_x\right)$ with $a_i=-i\sum_{E_n<0}\bra{n,{\bf k}}\partial/\partial k_i\ket{n,{\bf k}}$. For a system with Chern number $C_1=N$, on an open boundary there are $N$ channels of chiral Majorana edge states, described by the following Hamiltonian:
\bea
H_{\rm edge}^\uparrow=\int dkvk\sum_{a=1}^{|N|}\eta_{\uparrow ak}^\dagger\eta_{\uparrow ak}
\eea
with $k$ the momentum along the edge, and $\eta_{\uparrow ak}$ the edge state quasiparticle operator which is a superposition of electron and hole operators. 
$\eta_{\uparrow ak}$ satisfies the Majorana condition $\eta_{\uparrow ak}^\dagger =\eta_{\uparrow a,-k}$ so that the edge state fermion is its own anti-particle. In particular, $\eta_{\uparrow a,k=0}=\eta_{\uparrow a,k=0}^\dagger$ is a Majorana zero mode, carrying half of the degree of freedom of a complex fermion\cite{volovik1999,read2000}. For $N>0$ ($N<0$) $v$ is positive (negative) respectively, corresponding to left and right moving edge states. For simplicity we have set the velocity of the Majorana edge states to be the same. Generically the velocity can be different for different $a$ but it does not change the topological properties. According to Eq. (\ref{Tofh}) determined by time-reversal symmetry, it can be shown straightforwardly that $h^\downarrow$ and $h^{\uparrow}$ have opposite Chern number. Correspondingly the edge states of them consist of the same number of Majorana edge states with opposite chirality. For $C_1^\uparrow=-C_1^\downarrow=N>0$, the edge theory of the whole system is
\bea
H_{\rm edge}=\int dkvk\sum_{a=1}^{|N|}\left[\eta_{\uparrow ak}^\dagger\eta_{\uparrow ak}-\eta_{\downarrow ak}^\dagger\eta_{\downarrow ak}\right]\label{Hedge}
\eea
Since $\eta_{\uparrow(\downarrow)ak}$ is a superposition of $c_{\uparrow(\downarrow)}$, the action of $R$ symmetry on $\eta_{\uparrow(\downarrow)ak}$ is $R^{-1}\eta_{\uparrow ak}R=\eta_{\uparrow ak},~R^{-1}\eta_{\downarrow ak}R=-\eta_{\downarrow ak}$. Consequently, no mixing term such as $\eta_{\uparrow ak}^\dagger \eta_{\downarrow ak'}$ is allowed in the Hamiltonian $H_{\rm edge}$ if we consider generic perturbations preserving the $R$ symmetry.

In summary, from the analysis above we conclude that the 2D superconductors with $T$, $R$ symmetries are classified by integer, and the class labeled by integer $N$ has an edge theory of $|N|$ flavors of non-chiral Majorana fermions, which are protected to be gapless by the two symmetries.

\noindent{\it $Z_8$ Classification of interacting theory}. Now we consider the effect of electron interaction in this system. With interaction the terms mixing spin up and down electrons are allowed by the $R$ symmetry, such as a pair hopping term $H_{ph}=\sum_{ijkl}\left(t_{ijkl}c_{i\uparrow}^\dagger c_{j\uparrow}c_{k\downarrow}c_{l\downarrow}+h.c.\right)$ with $h.c.$ denotes the Hermitian conjugate of the first term. The time-reversal symmetry leads to some requirements on the matrix elements $t_{ijkl}=t_{klij}$. Other terms such as $c^\dagger_{i\uparrow}c^\dagger_{j\uparrow}c^\dagger_{k\downarrow}c^\dagger_{k\downarrow}$ can also be considered. To see the consequence of such interaction terms on the $\mathbb{Z}$ topological invariant, it is most convenient to study the stability of the edge states described by the Hamiltonian (\ref{Hedge}) under interaction. In the following we will show that the Hamiltonian (\ref{Hedge}) with $N=8$ can be gapped by an interacting term without breaking the $T,R$ symmetries, so that the classification of the bulk topological states can at most be $\mathbb{Z}_8$.

This conclusion is obtained based on the results of Ref. \cite{fidkowski2010} in $(1+1)$-dimensions. It was shown explicitly in Ref. \cite{fidkowski2010} that the following interacting term induces a mass to $8$ channels of free Majorana fermions given by Eq. (\ref{Hedge}):
\bea
H_{\rm int}&=&H_{1234}+H_{5678}+H_{1256}+H_{3478}+H_{3456}+H_{1278}\nonumber\\
& &-H_{2367}-H_{1458}-H_{2358}-H_{1467}\nonumber\\
& &+H_{1357}+H_{2468}-H_{1368}-H_{2457}\nonumber\\
H_{abcd}&=&\int dxM\epsilon^{abcdefgh}\eta_{e\uparrow}\eta_{f\uparrow}\eta_{g\downarrow}\eta_{h\downarrow}\label{Hint}
\eea
For $8$ Majorana fermions with the same velocity, there is an $SO(8)$ symmetry between the Majorana fermions generated by the operators $J_{ab}=\frac1{4i}\left(\left[\eta_{a\uparrow},\eta_{b\uparrow}\right]+\left[\eta_{a\downarrow},\eta_{b\downarrow}\right]\right)$, and the Majorana fermion operators $\eta_a$ form vector representation of this $SO(8)$. The interaction term given above breaks the $SO(8)$ symmetry to $SO(7)$, but the $SO(7)$ subgroup is not the one which preserves some vector $n_a$ of $SO(8)$, but the one which preserves some spinor $\psi_\alpha$ of $SO(8)$. In the supplementary material we provide an explicit explanation why the interaction term (\ref{Hint}) leads to such a nonconventional symmetry breaking.

It is straightforward to verify that $H_{\rm int}$ defined in Eq. (\ref{Hint}) preserves both $T$ and $R$ symmetries. Thus the existence of such a mass term suggests that the quadratic Hamiltonian (\ref{Hbulk}) with Chern number $N=8$ for spin up electrons is topologically trivial, since its edge states are not topologically robust when interaction is considered. For $N>8$ one can add the same mass term for any $8$ of the Majorana fermions, leading to the conclusion that the Hamiltonians with Chern number $N$ and $N-8$ have topologically equivalent edge states. If the edge states for $N<8$ is stable, one is lead to the conclusion that the topological classification of $T,R$ invariant topological superconductors is $\mathbb{Z}_8$ rather than $\mathbb{Z}$.

\noindent{\it Edge theory and the relation to 1d BDI class} Now we present further evidences that the systems with Chern number $N=0,1,...,7$ are indeed topologically distinct from each other. Consider to add the following spatially inhomogeneous mass term to the edge states:
\bea
H_m=\int dx\sum_{a=1}^{|N|}im(x)\eta_{\uparrow a}(x)\eta_{\downarrow a}(x)\label{TRbreakingmass}
\eea
with $m(x)\in\mathbb{R}$ real. Physically, such a mass term can be obtained by coupling the edge states of the TSC with an $s$-wave superconductor by superconducting proximity effect. In the lattice Hamiltonian (\ref{Hbulk}), the $s$-wave pairing term has the form of $H_\Delta=\sum_i\left(\Delta c_{i\uparrow}^\dagger c_{i\downarrow}^\dagger +h.c.\right)$, which pairs each fermion $c_{i\sigma}$ with its time-reversal partner $T^{-1}c_{i\sigma}T=\sigma c_{i-\sigma}$. Such a property of $s$-wave pairing is generic and basis independent\cite{wang2011}, so that in the basis of low quasiparticles the pairing term has the form of (\ref{TRbreakingmass}). The mass term $m(x)$ is proportional to the imaginary part ${\rm Im}\Delta$ of the $s$-wave pairing, since the real part ${\rm Re}\Delta$ preserves time-reversal symmetry and does not induce a mass of the edge states.

\begin{figure}
\includegraphics[width=7cm]{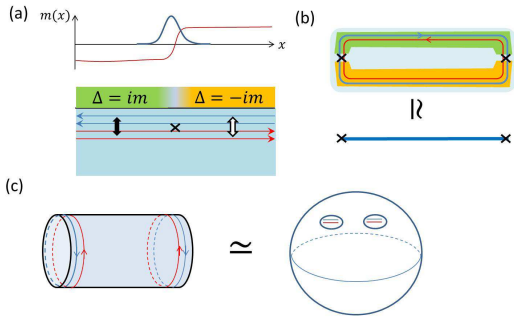}
\caption{(a) Illustration of the T-breaking mass domain wall along the edge of the 2d TSC. The mass is induced by a proximity effect of the TSC with an $s$-wave superconductor, and the mass domain wall corresponds to a Josephson junction between two such $s$-wave superconductors (see text). (b) Illustration of a strip of 2d TSC with the edge states on the two edges gapped by opposite mass terms $m$ and $-m$, which is topologically equivalent to a 1d TSC in the BDI class, with Majorana zero modes at the end. (c) Illustration that a sphere with two topological defects is equivalent to a cylinder with periodic boundary condition. } \label{fig1}
\end{figure}

In particular, consider the configuration of $m(x)$ with a domain wall where $m(x)$ changes its sign, as shown in Fig. \ref{fig1} (a). This configuration corresponds to a Josephson junction between two $s$-wave superconductors in proximity with the TSC, as shown in Fig. \ref{fig1} (a). As is well-known, such a mass term domain wall leads to Majorana zero modes localized on the domain wall\cite{jackiw1976}. For Chern number $N$ there are $|N|$ zero modes on the domain wall. As is known for other symmetry classes, such as the quantum spin Hall state with T-breaking domain wall on the edge\cite{qi2008}, such zero modes on the domain wall of a symmetry breaking mass term can carry fractional quantum numbers, in which case it can be used as a probe of the topological state even if electron interaction is considered. For example, in the case of quantum spin Hall insulator the T-breaking mass domain wall on the edge traps a fractional charge\cite{qi2008}. In the current system, to see if the zero modes carry fractional quantum numbers, one can first analyze the symmetry of the system with the domain wall. The mass term $H_m$ in Eq. (\ref{TRbreakingmass}) breaks both $T$ and $R$ symmetries, but preserves the combined symmetry $\tilde{T}=T\circ R$. From the definition of $T$ and $R$ in Eqs. (\ref{Tdef}) and (\ref{Rdef}) one can see that $\tilde{T}^{-1}c_{i\uparrow}\tilde{T}=c_{i\downarrow},~\tilde{T}^{-1}c_{i\downarrow}\tilde{T}=c_{i\uparrow}$, so that $\tilde{T}^2=1$. Thus $\tilde{T}$ can be considered as the time-reversal symmetry for spinless fermions, which is the symmetry defining the BDI symmetry class\cite{zirnbauer1996,altland1997}. The domain wall zero modes are thus topologically equivalent to the edge zero modes of a 1d topological superconductor in the BDI class\cite{fidkowski2010}. The relation between the domain wall zero modes and the edge state of 1d TSC can be seen most explicitly in the geometry shown in Fig. \ref{fig1} (b). A strip of 2d TSC with the upper and lower surface states gapped by opposite mass $m$ and $-m$ can be considered as a 1d TSC in the BDI class. With open boundary in the horizontal direction, the end of the 1d system becomes a mass domain wall between $m$ and $-m$ which has Majorana zero modes if the topological invariant is $N\neq 0$ mod $8$. In Ref. \cite{fidkowski2011,turner2011,chen2011} it was shown that the edge zero modes of a 1D system are classified by projective representations of the symmetry group, or equivalently, the second cohomology of the symmetry group. Here the symmetries of the system are $\tilde{T}$ and the fermion number parity $F$ which is always a symmetry for fermion systems. $\tilde{T}$ and $F$ commutes with each other, and form the symmetry group of $\mathbb{Z}_2\times\mathbb{Z}_2$. There are $8$ distinct projective representations of this group which corresponds to $N=0,1,...,7$ number of Majorana zero modes. Therefore from the results on 1D BDI class one can conclude that the domain wall zero modes for $N=1,2,..,7$ is stable, so that the corresponding bulk states are topologically distinct.


It is helpful to make more comments on the relation of the $N=8$ edge state problem with the $(1+1)$-d topological superconductor studied in Ref. \cite{fidkowski2010}. 
We start by considering more general mass terms in the edge theory which preserves the $\tilde{T}$ symmetry. A quadratic mass term is allowed in the edge theory in the form of $H_{m}=i\int dx A_{ab}\eta_{a\uparrow}(x)\eta_{b\downarrow}(x)$. The requirement of Hermitivity of the Hamiltonian and $\tilde{T}$ symmetry requires the condition $A_{ab}=A_{ba}\in\mathbb{R}$, {\it i.e.} the matrix $A$ is real and symmetric. Such a massive Majorana fermion can be viewed as a $(1+1)$-d topological superconductor in BDI class studied in Ref. \cite{fidkowski2010}, with a topological invariant $N={\rm Ind}(A)$ defined as the number of negative eigenvalues of mass matrix $A$. If only quadratic mass terms $H_m$ are considered, the topological classification would be $\mathbb{Z}$ since some eigenvalues of $A_{ab}$ must vanish at the transition between states with different index $N$, leading to the topological phase transition. However it is shown in Ref. \cite{fidkowski2010} that a quartic mass term $H_{\rm int}$ given in Eq. (\ref{Hint}) is possible for $N=8$, such that it is possible to have a cross over between two $A_{ab}$'s with index different by $8$. For example, consider the mass term
\bea
H(\theta)=im\cos\theta\int dx\sum_{a=1}^8\eta_{a\uparrow}\eta_{a\downarrow}+\sin\theta H_{\rm int}
\eea
For $m>0$, the Hamiltonian at $\theta=0,\pi$ has index $0$ and $8$, respectively. By adding the interaction term $H_{\rm int}$, one obtains an interpolation between $H(0)$ and $H(\pi)$ without closing the gap. On the edge of $(2+1)$-d topological superconductor, the additional $R$ symmetry requires the quadratic mass term $A_{ab}$ to vanish. Thus the edge state theory can be viewed as the topological phase transition theory between different topological states in the $(1+1)$-d BDI class. Due to the presence of the quartic mass term $H_{\rm int}$, there are only $8$ distinct phases in $(1+1)$-d BDI class. Thus there are $8$ distinct phase transition theories at presence of the $R$ symmetry, which becomes the robust topological edge states in the $(2+1)$-d case classified by $\mathbb{Z}_8$.

\noindent{\it Topological defect in the bulk}. Another important topological property of the topological superconductors in $D'$ class is the topological defects in the bulk. Generically, superconductors have vortices as point-like defects. If we restrict the discussion to $T,R$ invariant defects, the only vortices preserving the symmetries are those with opposite vorticity for spin up and down components. Majorana zero modes can appear in the vortex core. 
To understand the zero modes in the vortex core, one can relate the vortex zero modes to the edge states. As shown in Fig. \ref{fig1} (c), a sphere with two holes is topologically equivalent to a cylinder with two edges. The edge state on one edge of the cylinder is described by the edge Hamiltonian (\ref{Hedge}) compactified on a finite size circle: $H_{\rm edge}=\sum_{k,a}vk\sum_{a=1}^{|N|}\left[\eta_{\uparrow ak}^\dagger\eta_{\uparrow ak}-\eta_{\downarrow ak}^\dagger\eta_{\downarrow ak}\right]$. Their are two possible boundary conditions for the Majorana fermions on the cylinder, periodic or anti-periodic. When the cylinder is deformed to the sphere with two holes, the periodic and anti-periodic boundary conditions are mapped to the sphere with and without a vortex in each hole, respectively. For periodic boundary condition, the wavevector $k$ takes values of $k=2\pi n/L,~n\in\mathbb{Z}$ with $L$ the perimeter to the edge. When the perimeter $L$ shrinks, the states with nonzero $k$ has higher and higher energy $\propto 1/L$, and the only important low energy modes are the Majorana zero modes $\eta_{\uparrow a0},\eta_{\downarrow a0}$ with $k=0$. The coupling term between different spins such as $im_{ab}\eta_{\uparrow a0}\eta_{\downarrow b0}$ is forbidden by the $T$ and $R$ symmetries. However, different from the case of one-dimensional edge states, the zero modes of the same spin can be coupled to open a gap. For example if there are two zero modes $\eta_{\uparrow(\downarrow) a0},~a=1,2$, a mass term $im\left(\eta_{\uparrow10}\eta_{\uparrow 20}-\eta_{\downarrow 10}\eta_{\downarrow 20}\right)$ can be added to completely gap out the vortex zero modes. Therefore the vortex zero mode is classified by $\mathbb{Z}_2$ although the bulk topological invariant is $\mathbb{Z}_8$. This is similar to the vortices in the $D$ class which is also classified by $\mathbb{Z}_2$. When there are odd pairs of Majorana zero modes in the vortex core, no mass term can be introduced, and at least one pair of Majorana zero mode remains stable. Under time-reversal $T$ the single pair of Majorana zero mode forms a Kramers doublet, in the same way as the vortices in $DIII$ class with only $T$ symmetry\cite{qi2009b}.


\noindent{\it Conclusion and discussion}. In summary we have demonstrated that in $(2+1)$-dimensions there is a class $D'$ of TSC defined by two discrete symmetries which has a $Z_8$ topological classification. We studied the edge and bulk topological defects of the $D'$ TSC. The $(2+1)$-d $D'$ TSC is related to the $(1+1)$-d $BDI$ class of TSC with symmetry $\tilde{T}$ by adding one spatial direction and one additional discrete symmetry $T$. Such a relation between TSM in different dimensions applies to more generic cases. In Kitaev's ``periodic table" in Ref. \cite{kitaev2009} it is shown that the topological classification remains the same if the spatial dimension is increased from $d$ to $d+1$ while adding one symmetry $T$ satisfying $T^2=-1$. Our result demonstrates that such a relation between TSM in different dimensions with different symmetry can be generalized to interacting systems. The physical reason of such a correspondence has been illustrated in Fig. (\ref{fig1}) (b). Consider a $d+1$-dimensional system with one of the dimensions (the vertical direction in Fig. \ref{fig1} (b)) small compared to other dimensions. When the added symmetry ($T$ in the case of $D'$ class) in the $d+1$ dimensional system is broken on the boundary but preserved in the bulk, generically the surface states are gapped and such a $d+1$-d system can be viewed as a $d$-dimensional TSM in the known symmetry class. However there are always two ways to gap the surface states, which corresponds to doing a $T$ transformation to the boundary symmetry breaking terms. A domain wall forms between the two regions with opposite symmetry breaking mass terms, as is shown in Fig. \ref{fig1} (b). If the $d$-dimensional TSM obtained in this way is topological nontrivial, the domain wall must carry zero modes which are topologically equivalent to the boundary states of the $d$-dimensional TSM.

We would like to emphasize an important difference between the cases of interacting and non-interacting systems. For non-interacting TSM, adding a symmetry with $T^2=-1$ has the same effect as removing a symmetry with $\tilde{T}^2=-1$\cite{kitaev2009}. In the example of 1d BDI class, if one removes the $\tilde{T}$ symmetry one obtains the $D$ class (generic superconductors). Instead, adding the symmetry $T$ leads to the $D'$ class we discussed. Both $D$ and $D'$ class TSC are classified by integer in the non-interacting classification, but they are distinct in interacting systems. This is consistent with the fact that the edge domain wall construction only works when a discrete symmetry is added, rather than removed. More systematical discussion on this dimensional reduction approach to TSM is reserved for future works.

I acknowledge helpful discussions with Lukas Fidkowski and Alexei Kitaev. This work is supported by the Packard Foundation. Recently I became aware of several independent approaches to similar 2d topological states classified by $\mathbb{Z}_8$\cite{yao2012,ryu2012,gu2012b}.

\bibliography{TI}

\begin{appendix}
\begin{widetext}
\newpage

 \renewcommand{\theequation}{A-\arabic{equation}}
  \setcounter{equation}{0}  

\section*{Appendix: Further analysis on the interaction Hamiltonian (\ref{Hint})}

In this appendix we provide an explicit explanation on the symmetry property of the interaction term (\ref{Hint}). The analysis is already presented in Ref. \cite{fidkowski2010}, but we would like to provide an alternative illustration that is hopefully simpler and more explicit.

To see the symmetry breaking induced by the interaction term (\ref{Hint}), we introduce the Clifford algebra ${\rm Cliff}(8,0)$ with the generators $\Gamma_a,a=1,2,...,8$ satisfying
\begin{eqnarray}
\left\{\Gamma_a,\Gamma_b\right\}=2\delta_{ab}\label{Clifford}
\end{eqnarray}
$\Gamma_a$ can be represented by Hermitian matrices. If we don't require the representation of Clifford algebra to be real, the minimal dimension of faithful representation for the algebra ${\rm Cliff}(2n,0)$ is $2^n$. If we require the representation to be real $\Gamma_a=\Gamma_a^*$, usually the minimal dimension is larger than $2^n$. The special property of ${\rm Cliff}(8,0)$ is that the faithful representation with minimal dimension $2^4=16$ is also real. To obtain an explicit understanding, one can consider the following matrix representation of $\Gamma_a$:
\begin{eqnarray}
\Gamma_1&=&\nu_y\sigma_y\tau_0\mu_0\nonumber\\
\Gamma_2&=&-\nu_y\sigma_x\tau_0\mu_y\nonumber\\
\Gamma_3&=&\nu_y\sigma_x\tau_y\mu_z\nonumber\\
\Gamma_4&=&\nu_y\sigma_x\tau_y\mu_x\nonumber\\
\Gamma_5&=&-\nu_y\sigma_z\tau_x\mu_y\nonumber\\
\Gamma_6&=&\nu_y\sigma_z\tau_y\mu_0\nonumber\\
\Gamma_7&=&\nu_y\sigma_z\tau_z\mu_y\nonumber\\
\Gamma_8&=&\nu_x\sigma_0\tau_0\mu_0\label{repofGamma}
\end{eqnarray}
Here $\nu_i,\sigma_i,\tau_i,\mu_i$ are matrices in four independent $2\times 2$ spaces, which are defined as the Pauli matrices for $i=x,y,z$, and $2\times 2$ identity matrices for $i=0$. The multiplication in the equation above shall be understood as direct product, so that $\Gamma^a$ are $16\time 16$ matrices. It can be directly verified that this representation of $\Gamma^a$'s is real, and satisfies the Clifford algebra defined in Eq. (\ref{Clifford}).

We define the generators $T_{ab}$ of vector representation of $SO(8)$ as
\begin{eqnarray}
\left[T_{ab}\right]^{cd}=-i\left(\delta_{a}^c\delta_{b}^d-\delta_a^d\delta_b^c\right)
\end{eqnarray}
A generic $8\times 8$ orthogonal matrix is written as $O=\exp\left[iT_{ab}\theta^{ab}\right]$. The spinor representation of $SO(8)$ is generated by the commutators $\Gamma_{ab}=\left[\Gamma_a,\Gamma_b\right]/4i$. The elements of the spinor representation are $U=\exp\left[i\Gamma_{ab}\theta^{ab}\right]$. $\Gamma^a$ carries the vector representation of $SO(8)$ under the similarity transformation defined by $\Gamma^a\rightarrow U^{-1}\Gamma^aU\equiv O_{ab}\Gamma_b$. In the choice of representation (\ref{repofGamma}), all $\Gamma_{ab}$'s have the form of $\Gamma_{ab}=\nu_0...$ or $\Gamma_{ab}=\nu_z...$. Consequently all $\Gamma_{ab}$ are block diagonal with two $8\times 8$ blocks. In other words, the spinor representation is reducible. The two $8\times 8$ blocks are the spinor representation and the conjugate spinor representation of $SO(8)$. Moreover, because $\Gamma_a$ is real, $\Gamma_{ab}$ is purely imaginary and the rotation in the spinor representation $U$ is real. If we denote
\begin{eqnarray}
\Gamma_{ab}\equiv\left(\begin{array}{cc}\gamma_{ab}^+&0\\0&\gamma_{ab}^-\end{array}\right)
\end{eqnarray}
one can define the mappings from vector representation to the spinor representations
\begin{eqnarray}
\varphi_+:~O&=&\exp\left[iT_{ab}\theta^{ab}\right]\longrightarrow U_+=\exp\left[i\gamma^+_{ab}\theta^{ab}\right],\nonumber\\
\varphi_-:~O&=&\exp\left[iT_{ab}\theta^{ab}\right]\longrightarrow U_-=\exp\left[i\gamma^-_{ab}\theta^{ab}\right]\label{isom}
\end{eqnarray}
Because $U_\pm$ are also real orthogonal matrices, the two maps $\varphi_\pm$ are {\it isomorphisms} between the vector representation and the spinor representations. This is the so-called triality property of $SO(8)$.

Now we consider $8$ Majorana fermion operators $\eta_a$. The Hilbert space of such as system is $16$ dimensional, the same as that of $4$ complex fermions. Since $\eta_a$ satisfies the Clifford algebra in Eq. (\ref{Clifford}), we can take a basis in the Hilbert space so that the matrix representation of operators $\eta_a$ is $\Gamma_a$. The time-reversal symmetry $\tilde{T}$ acts on $\eta_a$ as $\tilde{T}^{-1}\eta_a\tilde{T}=\eta_a$. Therefore in the representation given in Eq. (\ref{repofGamma}) one can take the time-reversal symmetry to be $\tilde{T}=K$ which is the complex conjugation. Now we want to define a time-reversal invariant Hamiltonian in this Hilbert space with a unique ground state. The simplest choice is a projection operator:
\begin{eqnarray}
H_{\rm proj}=\left(\begin{array}{cccc}0&&&\\&0&&\\&&..&\\&&&-1\end{array}\right)_{16\times 16}\label{Hproj}
\end{eqnarray}
which has all entries vanishing except the lower right corner. Such a projector Hamiltonian breaks the $SO(8)$ of the lower $8\times 8$ block to $SO(7)$. However, this $SO(7)$ is the $SO(7)$ which rotates the first $7$ columns of the lower $8\times 8$ block. This is not the usual $SO(7)$ subgroup of $SO(8)$ which preserves a given vector. Instead, it's the subgroup which preserves a given spinor $(0,0,....,1)$.

We would like to express such a Hamiltonian in the Majorana fermion operators $\eta_a$, {\it i.e.}, $\Gamma_a$ in this representation. In general, a complete basis of $16\times 16$ real matrices can be obtained by multiplying different $\Gamma_a$'s. The set of $\left\{1,\Gamma_a,\Gamma_a\Gamma_b,\Gamma_a\Gamma_b\Gamma_c,....,\Gamma_1\Gamma_2....\Gamma_8\right\}$ (with all $a, b, c...$ in each term different from each other) contains exactly
\begin{eqnarray}
\sum_{n=0}^8\left(\begin{array}{c}8\\n\end{array}\right)=2^8\nonumber
\end{eqnarray}
independent real matrices. Because $\Gamma^a$ are all off-diagonal, the multiplication of even $\Gamma_a$'s are non-vanishing in the diagonal $8\times 8$ blocks and those of odd $\Gamma_a$'s are non-vanishing in the off-diagonal $8\times 8$ blocks. In this set, the matrices with nonzero overlap with $H_{\rm proj}$ are those which are both symmetric and block-diagonal. Beside the identity $1$, there are only two set matrices satisfying these two conditions: $\Gamma_a\Gamma_b\Gamma_c\Gamma_d$ and $F=\Gamma_1\Gamma_2....\Gamma_8$.
In the representation (\ref{repofGamma}) $F=-\nu_z$. Physically $F$ is the fermion number parity. Since the Hamiltonian (\ref{Hproj}) vanishes in the upper $8\times 8$ block, we can focus on the basis of the lower $8\times 8$ block obtained by the projection:
\begin{eqnarray}
\gamma_{abcd}=\frac12(1+F)\frac1{24}\epsilon^{abcdefgh}\Gamma_e\Gamma_f\Gamma_g\Gamma_h
\end{eqnarray}
The antisymmetric tensor $\epsilon^{abcdefgh}$ is introduced to extract only the components of $\Gamma_a\Gamma_b\Gamma_c\Gamma_d$ with $a,b,c,d$ all different from each other. $\gamma_{abcd}$ satisfies the orthogonality condition
\begin{eqnarray}
\frac18{\rm Tr}\left[\gamma_{abcd}\gamma_{efgh}\right]=\frac{1+F}{2}\left[\frac1{24}\epsilon^{abcdijkl}\epsilon_{ijklefgh}+\epsilon_{abcdefgh}\right]
\end{eqnarray}
For example $\gamma_{1234}=\frac12(1+F)\Gamma_5\Gamma_6\Gamma_7\Gamma_8=\frac12\left(\Gamma_1\Gamma_2\Gamma_3\Gamma_4+\Gamma_5\Gamma_6\Gamma_7\Gamma_8\right)$, which is orthogonal to other $\gamma_{abcd}$ except $\gamma_{5678}=\gamma_{1234}$. The number of independent $\gamma_{abcd}$ is therefore $\frac{8!}{2\times 4!}=35$ which is consistent with the number of symmetric traceless $8\times 8$ matrices $35=\frac{8\times 9}2-1$. Using the orthogonality condition one can expand the Hamiltonian to
\begin{eqnarray}
H_{\rm proj}=\frac{1+F}{16}{\rm tr}\left[\frac{1+F}2H_{\rm proj}\right]+{\sum_{abcd}}'\frac18{\rm tr}\left[\gamma_{abcd}H_{\rm proj}\right]\cdot \gamma_{abcd}
\end{eqnarray}
with the first term taking care of the trace part and the second term an expansion of the traceless part in $\gamma_{abcd}$. Here the sum $\sum_{abcd}'$ stands for the sum over the $35$ independent groups of $abcd$. For the particular Hamiltoinan $H_{\rm proj}$, we have
\begin{eqnarray}
{\rm tr}\left[\gamma_{abcd}H_{\rm proj}\right]=-\left[\gamma_{abcd}\right]_{16,16}
\end{eqnarray}
which is the last diagonal component of $\gamma_{abcd}$. In the representation given by Eq. (\ref{repofGamma}), $\gamma_{abcd}$ is also a direct product of Pauli matrices or identity. Consequently the $16,16$ component is only nonzero if $\gamma_{abcd}=\nu_i\sigma_j\tau_k\mu_l$ with all $i,j,k,l=0$ or $z$. Therefore there are only $7=2^3-1$ terms with nonzero contributions. They can be explicitly found as
\begin{eqnarray}
\gamma_{1234}&=&-\frac{\nu_0-\nu_z}2\sigma_z\tau_0\mu_0,~\gamma_{1256}=\frac{\nu_0-\nu_z}2\sigma_z\tau_z\mu_0,~\gamma_{1278}=-\frac{\nu_0-\nu_z}2\sigma_0\tau_z\mu_0\nonumber\\
\gamma_{1357}&=&\frac{\nu_0-\nu_z}2\sigma_z\tau_0\mu_z,~\gamma_{1368}=\frac{\nu_0-\nu_z}2\sigma_0\tau_0\mu_z,~\gamma_{1458}=-\frac{\nu_0-\nu_z}2\sigma_0\tau_z\mu_z,~
\gamma_{1467}=\frac{\nu_0-\nu_z}2\sigma_z\tau_z\mu_z
\end{eqnarray}
Therefore we obtain
\begin{eqnarray}
H_{\rm proj}=-\frac{1+F}{16}-\frac18\left[\gamma_{1234}+\gamma_{1256}+\gamma_{1278}+\gamma_{1357}-\gamma_{1368}-\gamma_{1458}-\gamma_{1467}\right]\label{Hproj2}
\end{eqnarray}
$\gamma_{abcd}$ can be written in product of $\Gamma_a$'s, which in the operator form are $\eta_a$'s. For example $\gamma_{1234}=\frac12\left(\eta_1\eta_2\eta_3\eta_4+\eta_5\eta_6\eta_7\eta_8\right)$. Compare Eq. (\ref{Hproj2}) with Eq. (\ref{Hint}) one finds that except for a trivial constant term, they are identical if we only take the zero modes $\eta_{a,k=0\uparrow(\downarrow)}$ in Eq. (\ref{Hint}). Consequently, we have demonstrated that the Hamiltonian (\ref{Hint}) leads to a unique ground state in the Hilbert space of the $8$ Majorana zero modes.

%

\end{widetext}
\end{appendix}

\end{document}